\documentclass[aps,prc,preprint,tightenlines,superscriptaddress,showpacs,%
amssymb,byrevtex,nofootinbib]{revtex4}
\usepackage[dvips]{graphicx,color}
\usepackage{dcolumn}
\usepackage{epsfig}
\usepackage{bm}

\begin{document}
\title{Production of Bound States with Hidden Charm at J-PARC and JLab }
\author{ Jia-Jun Wu}
\affiliation{Physics Division, Argonne National Laboratory, Argonne, Illinois 60439, USA}
\author{T.-S. H. Lee}
\affiliation{Physics Division, Argonne National Laboratory, Argonne, Illinois 60439, USA}
\begin{abstract}
The cross sections of the production of bound states with hidden charm
in reactions induced by pions and photons are presented.
To facilitate the experimental efforts to determine the $J/\Psi$-N
interactions, we present results for
 $J/\Psi$ production on the deuteron.
\end{abstract}

\maketitle

\section{introduction}

In a recent paper\cite{wulee12}, we presented predictions of
photo-production of bound states $[^3He]_{J/\Psi}$ on $^4He$ target and
$[q^6]_{J/\Psi}$ on $^3He$ target. In this work we apply the
same approach to
predict the production cross sections for these bound states with pion beams.

Our predictions depend on a potential
 model of $J/\Psi$-N interaction $v_{J/\Psi N,J/\Psi N}$. All theoretical
calculations of $v_{J/\Psi N,J/\Psi N}$, based on the
 effective field theory method\cite{pesk79,luke92,brodsky-1,russia},
the Pomeron-quark coupling model\cite{brodsky90},
and the Lattice QCD\cite{lqcd}, give an
 attractive feature.
 However the resulting strength of
$v_{J/\Psi N,J/\Psi N}$ is rather uncertain.
Here we present results on the deuteron target
to facilitate the experimental tests of these models.

\section{Production of bound states $[^3He]_{J/\Psi}$ and $[q^6]_{J/\Psi}$}

\begin{figure}[ht]
\centering
\epsfig{file=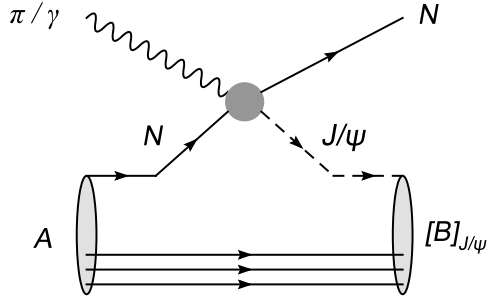, width=0.4\hsize}
\caption{The impulse approximation mechanism of
$\gamma/\pi + A \rightarrow N + [B]_{J/\Psi}$ reaction.
$A$ is a nucleus with mass number $A$ and $B$ could be a nucleus with mass
number $(A-1)$ or a $[q^{3(A-1)}]$ multi-quark cluster.
}
\label{fig:impulse}
\end{figure}

The calculations in Ref.\cite{wulee12}
 are based on the impulse approximation mechanism illustrated
in Fig.\ref{fig:impulse}. The same approach can be used  to perform
calculations with incident pions
by simply replacing
the $\gamma + N \rightarrow J/\Psi +N$ amplitude by the
$\pi+ N \rightarrow J/\Psi +N$ amplitude.
Following the approach of
Refs.\cite{wulee12} and \cite{brodsky-1}, we calculate
the  $\pi+ N \rightarrow J/\Psi +N$ amplitude
  from the $\rho$-exchange mechanism
calculated from the following Lagrangian
\begin{eqnarray}
L = L_{J/\Psi,\rho\pi} + L_{\rho NN}
\label{eq:larg}
\end{eqnarray}
with
\begin{eqnarray}
L_{J/\Psi,\rho\pi}&=& -\frac{g_{J/\Psi,\rho\pi}}{m_{J/\Psi}}
\epsilon^{\alpha\beta\mu\nu}
\partial_\alpha \phi_{J/\Psi,\beta}
\partial_\mu \vec{\rho}_\nu \cdot\vec{\phi}_{\pi}\,,
\label{eq:L-jrp} \\
L_{\rho NN}&=& \bar{\psi}_N[\gamma^\eta -
\frac{\kappa_\rho}{2m_N}\sigma^{\eta\delta}]\vec{\rho}_\eta\cdot
\frac{\vec{\tau}}{2}\psi_N\,,
\label{eq:L-rnn}
\end{eqnarray}
where $g_{J/\Psi,\rho\pi}=0.032$ is determined from the
width of $J/\Psi \rightarrow \rho + \pi$,
$g_{\rho NN}= 6.23$ and $\kappa_\rho =1.825$
are taken from a dynamical model\cite{sl96} of $\pi N$ scattering.

For the calculations of $\pi +^4He \rightarrow [^3He]_{J/\Psi}+N $, we need to
calculate the bound state wavefunction from a $^3He$-$J/\Psi$
potential $V_{3,J/\Psi}= \alpha_3 \frac{e^{-\mu r}}{r}$. Here we use
$\alpha_3=0.33$ and $\mu=257$ MeV determined\cite{wulee12} from
the Pomeron-quark coupling model of
Brodsky, Schmidt, and de Teramond\cite{brodsky90}.
For the calculations of $\pi +^3He \rightarrow [q^6]_{J/\Psi} +N$,
 the probability of finding a six-quark cluster $q^6$ in $^3He$
is determined by
using the  Compound Bag model\cite{fasano} of $NN$ interaction.
The relative wavefunction of $J/\Psi$-$q^6$ is
constrained by reproducing the $^3He$ charge form factor, as
 detailed in Ref.\cite{wulee12}.

\begin{figure}[ht]
\centering
\epsfig{file=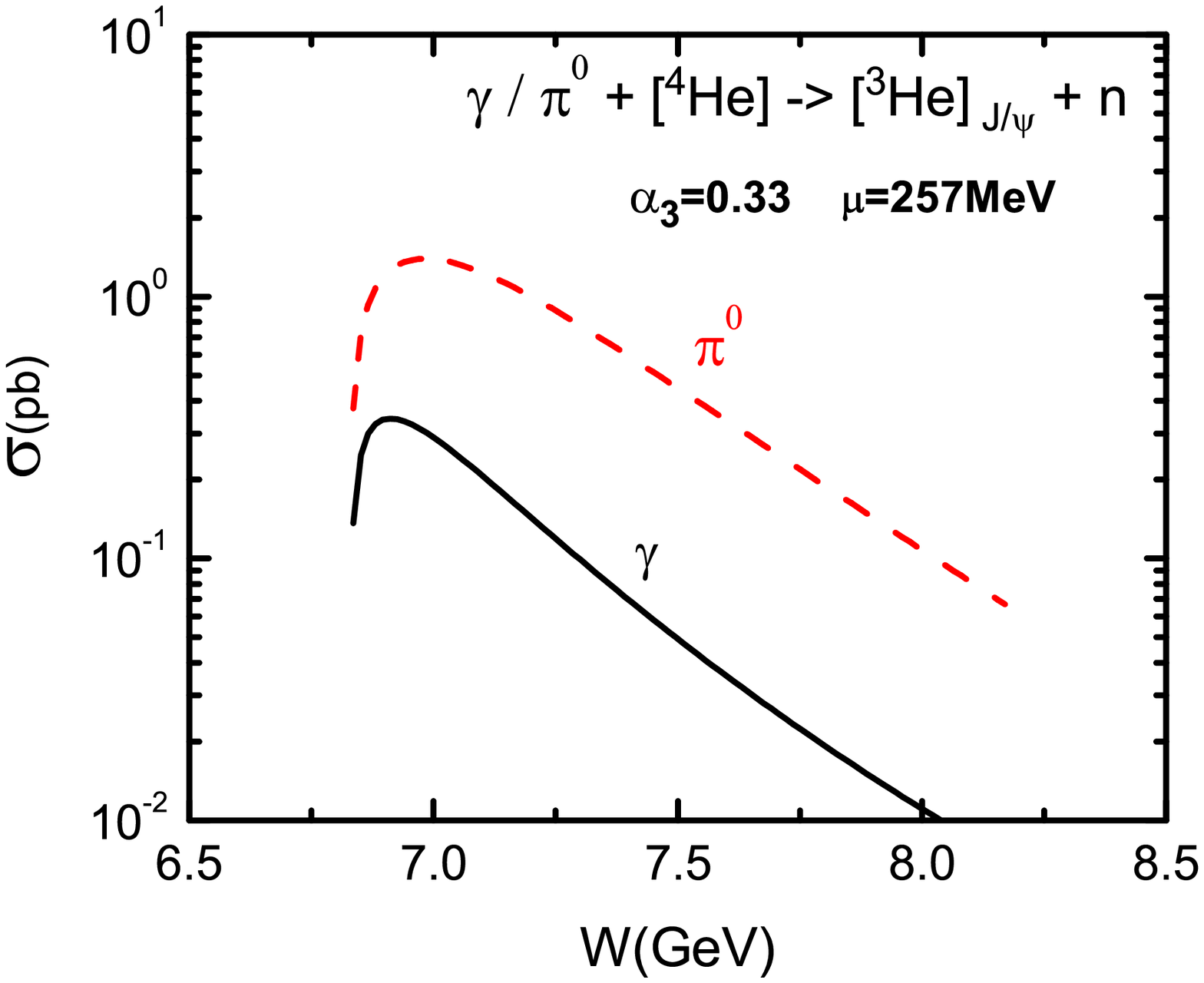, width=0.4\hsize}
\caption{Production cross sections of $\gamma/\pi +^4He \rightarrow
[^3He]_{J/\Psi} + n$.
}
\label{fig:he4}
\end{figure}

\begin{figure}[ht]
\centering
\epsfig{file=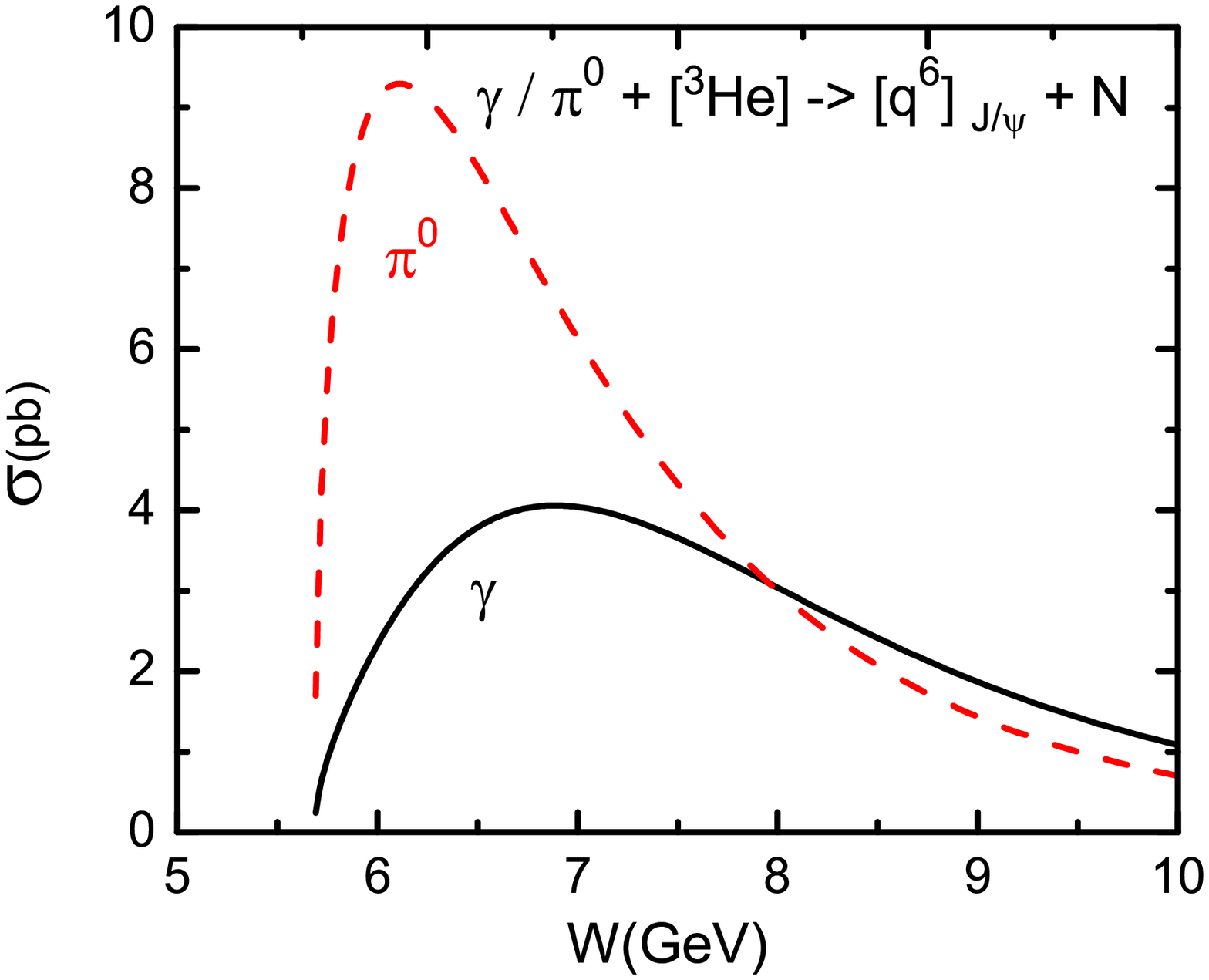, width=0.4\hsize}
\caption{ Production cross sections of $\gamma/\pi +^3He \rightarrow
[q^6]_{J/\Psi} + n$
}
\label{fig:he3}
\end{figure}

The results from pion and photon beams are compared
in Fig.\ref{fig:he4} for the $[^3He]_{J/\Psi}$ production on a $^4He$ target
 and in Fig.\ref{fig:he3} for $[q^6]_{J/\Psi}$ production
on a $^3He$ target. We see that the cross sections from pions
are about a factor of 2-3 larger than those from photons.
We also see that  the detections of these bound states with hidden charm
are favored at energies near the production threshold.

\section{Production on deuteron target}

\begin{figure}[ht]
\centering
\epsfig{file=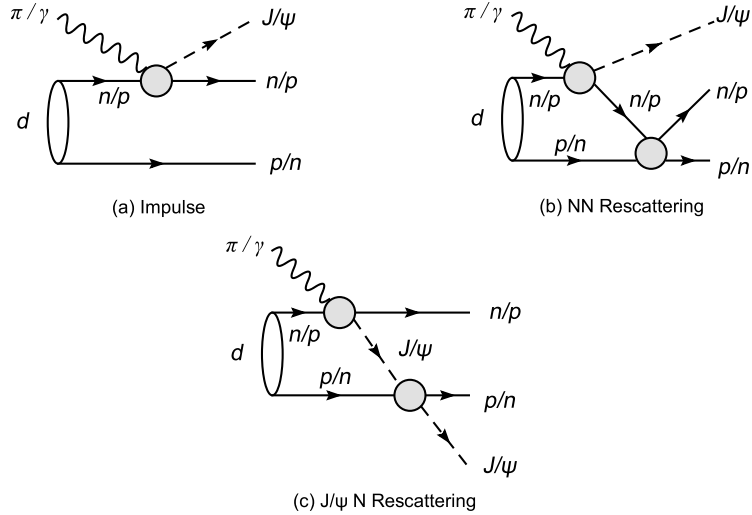, width=0.6\hsize}
\caption{The mechanisms of $\gamma + d \rightarrow J/\Psi + p + n$.
}
\label{fig:diagram}
\end{figure}

To facilitate the experimental determination of the
$J/\Psi$-N interaction, we make predictions of
the cross sections of $\gamma/\pi + d \rightarrow J/\Psi + n + p$.
In the impulse approximation, the amplitude of this process
is the coherent sum of the three mechanisms illustrated in
Fig.\ref{fig:diagram}.
The $\pi + N \rightarrow J/\Psi+ N$ amplitudes needed in the
calculations are computed from the $\rho$-exchange mechanism,
as described in section II.
The $\gamma + N \rightarrow J/\Psi+ N$ amplitude is taken from
Ref.\cite{wulee12}, $NN \rightarrow NN$ amplitudes are generated from
the Bonn potential, and $J/\Psi+ N \rightarrow J/\Psi+N $ amplitudes
are generated from a potential
$v_{J/\Psi N,J/\Psi N} = -\alpha \frac{e^{-\mu r}}{r}$.
With $\mu = 630$ MeV, the strength   $\alpha$ determines the
s-wave scattering lengths $a$. In presenting our results, we
use $a$ to indicate the strength
of the considered $J/\Psi$-N potential model.

\begin{figure}[ht]
\centering
\epsfig{file=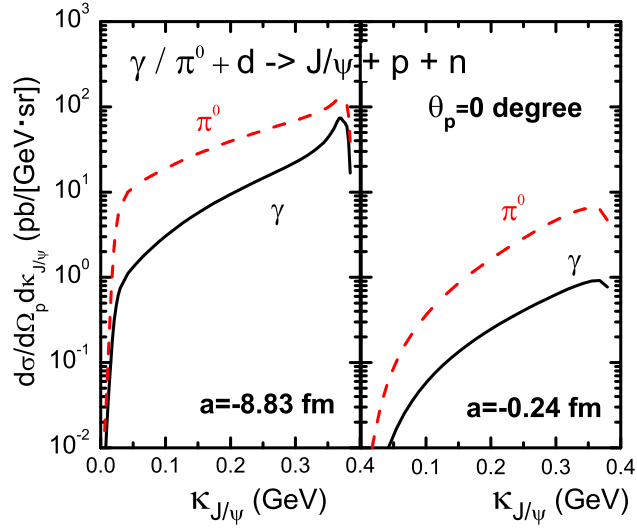, width=0.6\hsize}
\caption{The differential cross sections
of $\gamma/\pi + d \rightarrow J/\Psi + n + p$
}
\label{fig:alld}
\end{figure}

We find that the kinematics favoring the determination of
$v_{J/\Psi N,J/\Psi N}$ is in the region where the outgoing
proton is in the $\theta_p=0$ forward angle.
In Fig.\ref{fig:alld}, we compare the predicted differential cross section of the outgoing
proton at $\theta_p=0$, where
 $\kappa_{J/\Psi}$ denotes the relative momentum of the outgoing
$J/\Psi$-n  pair. We see that cross sections for pion beam are larger
than that for the photon beams in the low $\kappa_{J/\Psi}$
region where the $J/\Psi$-N relative motion is slow.
We also see that the predicted magnitudes depend on the
scattering length $a$ of the $J/\Psi$-N potential model.
The results for $a = -8.83$ fm(left) are about a factor of
10 larger than those for $a = -0.24$ fm(right).

\begin{figure}[ht]
\centering
\epsfig{file=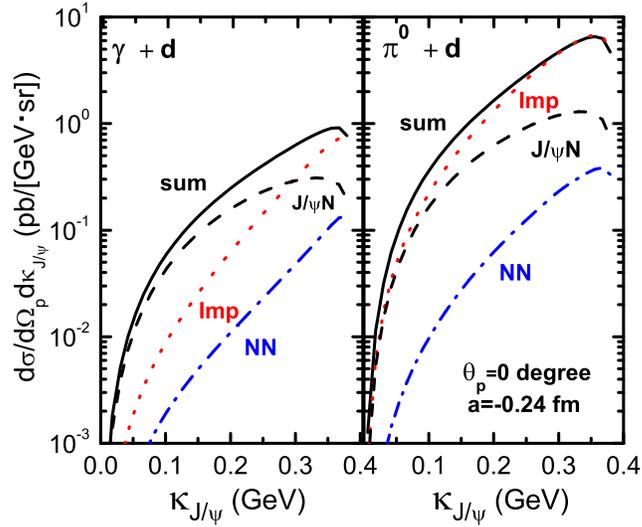, width=0.6\hsize}
\caption{ Relative importance between the contributions
from three mechanisms
illustrated in Fig.\ref{fig:diagram} to the differential cross sections
of $\gamma/\pi + d \rightarrow J/\Psi + n + p$.
Imp: Fig.4(a), $NN$: Fig.4(b), $J/\Psi N$: Fig.4(c).
}
\label{fig:diffd}
\end{figure}

In Fig.\ref{fig:diffd}, we show the relative importance between
the different mechanisms illustrated in Fig.\ref{fig:diagram}.
For the case with photon beams (left), the $J/\Psi$-N re-scattering term
(Fig.\ref{fig:diagram}(c))
dominants in the considered kinematic region. Thus the measured
cross section (solid curve) can be used to sensitively test
the considered $J/\Psi$-N potential models.
For the results with pion beams (right),  determinations of
 $J/\Psi$-N  interaction clearly need an accurate calculation of
the impulse term (Fig.\ref{fig:diagram}(a)), which is comparable
to the $J/\Psi$-N re-scattering term
(Fig.\ref{fig:diagram}(c)).

\section{discussions}
If the predicted bound states $[^3He]_{J/\Psi}$ and $[q^6]_{J/\Psi}$
can be detected, it will provide useful information
to understand the role of the gluon field in determining nuclear properties.
Thus the experiments on
 $\gamma/\pi +^4He (^3He) \rightarrow N + [^3He]_{J/\Psi} ([q^6]_{J/\Psi})$
will be very interesting to perform at J-PARC and JLab.
However, the data  can  be analyzed properly only when we have information to
determine the basic $J/\Psi$-N interactions. Our predictions on
the cross sections for $\gamma/\pi + d \rightarrow J/\Psi +n + p$
can facilitate the future  experimental efforts in this direction.

\clearpage

\begin{acknowledgments}
This work is supported by the U.S. Department of Energy, Office of Nuclear Physics Division,
under Contract No. DE-AC02-06CH11357.
This research used resources of the National Energy Research Scientific Computing Center,
which is supported by the Office of Science of the U.S. Department of Energy
under Contract No. DE-AC02-05CH11231, and resources provided on ``Fusion,''
a 320-node computing cluster operated by the Laboratory Computing Resource Center
at Argonne National Laboratory.
\end{acknowledgments}

\end{document}